\documentclass[12pt,a4paper,epsfig]{article}
\usepackage[latin1]{inputenc}
\usepackage{amsmath}	
\usepackage{amsfonts}
\usepackage{amssymb}
\usepackage{fullpage}
\usepackage{graphicx}
\usepackage{braket}
\usepackage{cite}
\usepackage{verbatim}
\usepackage[affil-it]{authblk}
\usepackage{titlesec}
\titlelabel{\thetitle.\ }
\usepackage{mathtools}
\usepackage{epstopdf}
\usepackage{bm}
\usepackage{hyperref}
\hypersetup{
    colorlinks=true,
    linkcolor=blue
}

\title{Geometric phase for neutrino propagation in magnetic field}

\author{{Sandeep Joshi\thanks{\textit{sjoshi@barc.gov.in}}} }  
\author{Sudhir R. Jain\thanks{\textit{srjain@barc.gov.in}}\footnote{Tel.: +912225593589}}
\affil{\mbox{Nuclear Physics Division, Bhabha Atomic Research Centre, Mumbai 400085, India}}

\date{}

\begin{document}
\maketitle

\begin{abstract}
The geometric phase for neutrinos propagating in an adiabatically varying magnetic field in matter is calculated. It is shown that for neutrino propagation in sufficiently large magnetic field the neutrino eigenstates develop a significant geometric phase. The geometric phase varies from 2$\pi$ for magnetic fields $\sim$ fraction of a micro gauss to $\pi$ for fields $\sim 10^7$ gauss or more. The variation of geometric phase with magnetic field parameters is shown and its phenomenological implications are discussed.
\end{abstract}

\section{Introduction}
The existence of non-zero neutrino mass is so far the only experimental proof \cite{ nu_mass} of physics beyond standard model. This calls for a minimal extension of standard model wherein a Dirac neutrino acquires a magnetic moment given by \cite{nu1}
\begin{equation}\label{eq:mu_nu}
\mu_\nu = 3.2\times10^{-19} \left(\frac{m_\nu}{1eV}\right) \mu_B,
\end{equation}
where $m_\nu$ is the neutino mass and $\mu_B$ is the Bohr Magneton, $\mu_B= 5.8 \times 10^{-15} MeV/G$.
Many theoretical considerations \cite{ babu,bell, nu5, rmp} which utilize physics beyond minimally extended standard model put much more stronger upper bound on the neutrino magnetic moments, as large as $\approx 10^{-14}\mu_B$ for a Dirac neutrino and $\approx 10^{-12}-10^{-10} \mu_B$ for a Majorana neutino.  Thus experimental observation of neutrino magnetic moment greater than Eq.\eqref{eq:mu_nu} would be a definite evidence of physics beyond minimally extended standard model. Also the results from the GEMMA experiment \cite{gemma} puts an upper bound on the neutrino magnetic moment at $2.9\times10^{-11}\mu_B$. 
The non-zero magnetic moment imparts non-trivial electromagnetic properties to the neutrino which opens up ``a window to new physics" \cite{rmp}.

	Inspired by the fact that the neutrinos have a non-zero magnetic moment, we envisage a physical situation where its effect could manifest in the presence of magnetic field. Such situations are encountered in astrophysical environments where neutrinos propagate through large distances in magnetic field in vacuum and in matter. As the neutrino propagates in a magnetic field, its magnetic moment will couple to the field leading to neutrino spin rotation, which may result in spin-flip transitions of neutrinos.

	For the case of neutrino spin precession in presence of a magnetic field in matter with constant density, the spin conversion probability is given by \cite{rmp}
\begin{equation}
P_{\nu_L \rightarrow \nu_R}(x)= \frac{(2\mu_\nu B_\perp)^2}{V^2+(2\mu_\nu B_\perp)^2} \sin^2 {\left(\frac{1}{2} \sqrt{V^2+(2\mu_\nu B_\perp)^2}x\right)},
\end{equation}
where $\mu_\nu$ is the neutrino magnetic moment, $B_\perp$ is the perpendicular component of the magnetic field and $V$ is the weak interaction potential for neutrinos propagating in matter. For a neutrino of flavor $l$, matter potential is given by $V_l= V_{CC}\delta_{le} + V_{NC}$,  $V_{CC}$ and $V_{NC}$ being charged-current and neutral current potentials respectively.

	There may also arise spin-flavor precession of neutrinos due to transition magnetic moments, which involves both spin flip and flavor conversion of the neutrino in presence of a magnetic field. If the magnetic field is twisting uniformly in the transverse plane of propagating neutrino, then the spin-flavor conversion probability in matter with constant density is given by \cite{smirnov, lim, ane}
\begin{equation}\label{prob}
P_{\nu_L \rightarrow \nu_R}(x)= \frac{(2\mu_\nu B_\perp)^2}{(V_m-\dot{\phi})^2+(2\mu_\nu B_\perp)^2} \sin^2 {\left(\frac{1}{2} \sqrt{(V_m-\dot{\phi})^2+(2\mu_\nu B_\perp)^2}x\right)},
\end{equation}
where $V_m= \sqrt{2}G_Fn^{eff}- \frac{\Delta{m}^2}{2E_\nu}\cos 2\theta$; $G_F$ is the Fermi's constant, $\Delta m^2$ is the mass squared difference $\Delta{m}^2 = m^2(\nu_{L})-m^2(\nu_{R})$, $\theta$ is the neutrino mixing angle,  $E_\nu$ is the neutrino energy, $n^{eff}$ is the effective concentration of particles interacting with neutrinos,
\begin{equation}\label{eq:neff}
n^{eff} = \begin{cases} (n_e-n_n), & \mbox{for } \nu_{eL}-\bar\nu_{\mu, \tau R} \\ 
						 (n_e-n_n/2), & \mbox{for } \nu_{eL}-\nu_{eR} \end{cases},
\end{equation}
$n_e$ and $n_n$ are the concentrations of electrons and neutrons respectively, and $\dot{\phi}$ is the precession frequency of the magnetic field.

	The spin-precession $\nu_{eL} \rightarrow \nu_{eR}$ in the magnetic field of the sun was considered by Cisneros\cite{cisneros} as a possible solution to account for the deficit of the active solar $\nu_e$ flux. Okun \textit{et al.} \cite{okun} included matter effects and showed that it results in the suppression of $\nu_{eL} \rightarrow \nu_{eR}$ transitions. It was soon realised \cite{akhmedov, lim, smirnov} that the spin-flavor precession of the neutrinos can lead to resonant conversion $\nu_e\rightarrow \bar{\nu}_{\mu,\tau}$. However, it  became eventually clear that the resonant spin flavor precession is not the correct solution of the solar neutrino problem\cite{barranco}.

	The geometric phase, on the other hand,  is a general property of quantum systems which arises if the Hamiltonian governing the system contains two or more time-dependent parameters. If the system is initially prepared in an eigenstate of the Hamiltonian then as the system undergoes an adiabatic evolution along a closed curve in the parameter space, the eigenstate develops a geometric phase in addition to the usual dynamical phase \cite{berry}. The geometric phase can be generalised for systems undergoing non-adiabatic, non-cyclic and non-unitary evolutions \cite{aharonov, samuel, mukunda} and has observable consequences in a wide variety of physical systems ranging from classical motion of Foucault pendulum to nuclear magnetic resonance studies \cite{anandan}. 
	
	The most common example where geometric phase arises is the spin-precession of a particle with an intrinsic magnetic moment in a magnetic field. The possibility of geometric phase associated with neutrino spin precession was explored in \cite{vidal, smirnov} in the context of solar neutrino problem. In \cite{smirnov} it was shown that the geometric phase may induce resonant spin conversion of neutrinos inside the sun. Also, there has been some interesting earlier work on geometric phase in neutrino oscillations where electromagnetic interactions of neutrinos are not included. Naumov\cite{naumov} calculated Berry phase for a three-flavor Dirac neutrino system as the neutrino propagation occurs in a medium whose density and element composition varies cyclically with distance.  He \textit{et al.}\cite{he} generalised \cite{naumov} and studied Berry phase in neutrino oscillations for both Dirac and Majorana neutrinos including active and active-sterile neutrino mixing and non-standard interactions.  The above papers \cite{naumov, he} claim that for non-trivial Berry phase to arise, the neutrino interactions with background matter must depend on two independent densities, and CP violating phase must be non-zero. Blasone \textit{et al.} \cite{blasone} studied Berry phase for the case of two and three flavor neutrino oscillations in vacuum. They showed that the cyclic time evolution of a neutrino flavor state produces an overall phase factor which, following Aharonov-Anandan prescription \cite{aharonov}, can be written as sum of two parts, a purely dynamical part and a part which depends only on the mixing angle and hence geometric in nature. Wang \textit{et al.} \cite{wang} generalised the results of \cite{blasone} for the case of non-cyclic time evolution. Mehta \cite{mehta} examined geometric phase for two neutrino flavor oscillations with CP conservation and showed that non-zero geometric phase appears not just at the amplitude level \cite{blasone, wang} but even at the level of probability and hence is directly observable. Syska \textit{et al.}\cite{syska} studied the geometric phase in $\pi^+$ decay and pointed out that geometric phase value of $\pi$ in neutrino oscillations is consistent with the mixing matrix parameters.

	Since it is known that the geometric phase does not have significant effect in causing the spin or spin-flavor transitions of neutrinos in the sun, we approach the problem from a more general perspective. We assume that the neutrinos are propagating  in an adiabatically varying arbitrary magnetic field in matter with constant density and, following Berry\cite{berry}, we explicitly calculate the geometric phase that arises in such situation. We then write transition probabilty $P({\nu_L \rightarrow \nu_R})$ in terms of geometric phase and analyse the values of geometric phase which lead to resonance.

\section{Geometric Phase}
We assume that the neutrino is propagating along the  z-direction, in a magnetic field rotating arbitrarily about the direction of motion of neutrino. The magnetic field vector can be written as
 \begin{equation}\label{eq:B}
 \textbf{B}(t) = B_0(\sin\theta\cos\phi, \sin\theta\sin\phi, \cos\theta),
 \end{equation}
 where $B_0$, $\theta$, and $\phi$ are adiabatically varying time dependent parameters.

The evolution equation describing the propagation of the two helicity components of a neutrino $\Ket\nu= (\nu_L, \nu_R)^T$ in magnetic field Eq.\eqref{eq:B} in matter is given by\\
\begin{equation}
i\frac{d}{dt}\Ket{\nu(t)}= H(t)\Ket{\nu(t)},
\end{equation}
where the Hamiltonian $H= H_0+ H_{wk}+ H_{em}$ ; $H_0$ is the vacuum Hamiltonian, $H_{wk}$ is the weak-interaction Hamiltonian for neutrino coupling with matter, and $H_{em}$ is the electromagnetic Hamiltonian for neutrino magnetic moment coupling with external magnetic field. We assume that CP is conserved and for simplicity neglect neutrino mass mixing, so that the flavor eigenstates coincide with the mass eigenstates.

	In the $(\nu_L, \nu_R)^T$ basis the total Hamiltonian, neglecting the terms proportional to unit matrix, can be written as \cite{smirnov}:
\begin{equation}
H= \begin{pmatrix}
\frac{V}{2}+\mu_{\nu}B_0\cos\theta & \mu_{\nu}B_0e^{-i\phi}\sin\theta \\ \\
\mu_{\nu}B_0e^{i\phi}\sin\theta	&	-\frac{V}{2}-\mu_{\nu}B_0\cos\theta
\end{pmatrix} ,
\end{equation}
where $V= \sqrt{2}G_Fn^{eff}- \frac{\Delta{m}^2}{2E}$ , $\Delta{m}^2 = m^2(\nu_{L})-m^2(\nu_{R})$, $E$ is the neutrino energy, $n^{eff}$ is the effective concentration of particles interacting with neutrinos given by Eq.\eqref{eq:neff}.

The instantaneous eignevalues and normalised eigenvectors of H are  \\
\begin{equation}
\lambda_\pm= \pm\sqrt{\left(\frac{V}{2}+\mu_{\nu}B_0\cos{\theta}\right)^2 + \left(\mu_{\nu}B_0\sin{\theta}\right)^2},
\end{equation}
\\
\begin{equation}
\Ket{+}= \frac{1}{N}\begin{pmatrix} \mu_{\nu}B_0\sin{\theta} \\ \\ -e^{i\phi}(\frac{V}{2}+\mu_{\nu}B_0\cos{\theta}-\lambda_+) \end{pmatrix},
 \end{equation},
 
 \begin{equation}  
\Ket{-}= \frac{1}{N}\begin{pmatrix} e^{-i\phi}(\frac{V}{2}+\mu_{\nu}B_0\cos{\theta}-\lambda_+) \\ \\ \mu_{\nu}B_0\sin{\theta}
\end{pmatrix} ,
\end{equation} 
\\ 
 where $\Ket{+}$ and $\Ket{-}$ corresponds to $\lambda_+$ and $\lambda_-$ respectively, and \\
 \begin{equation}
  N= \sqrt{\left(\frac{V}{2}+\mu_{\nu}B_0\cos{\theta}-\lambda_+\right)^2 + \left(\mu_{\nu}B_0\sin{\theta}\right)^2}.
 \end{equation} 
 
	 If the neutrino is initially in state $\Ket{+}$ at t=0, then as it moves along in the magnetic field it picks up a geometrical phase factor given by \cite{ berry}:
 \begin{equation} \label{eq:gamma}
\gamma_+ = i \oint_C \mathbf{dr} \cdot \Braket{+|\bm{\nabla}{+}},
 \end{equation} 
where the integral is over closed curve C in the parameter space.
	
	 For our case we have 
\begin{equation}
$$ \begin{flushleft}$$
\bm{\nabla}\Ket{+}= \frac{1}{B_0N}\begin{pmatrix}
 \mu_{\nu}B_0\cos{\theta}-f\mu_{\nu}B_0\sin{\theta} \\  \\
 e^{i\phi}[\mu_{\nu}B_0\sin{\theta}(1-\frac{V}{2\lambda_+})+f(\frac{V}{2}+\mu_{\nu}B_0\cos{\theta}-\lambda_+)]
\end{pmatrix}\hat{\theta} +$$\begin{center}$$
\frac{1}{NB_0\sin{\theta}} \begin{pmatrix}
0 \\ \\ -ie^{i\phi}(\frac{V}{2}+\mu_{\nu}B_0\cos{\theta}-\lambda_+)
\end{pmatrix}\hat{\phi} , $$\end{center}
\end{flushleft}$$
\end{equation} 
where \begin{equation}  f= \lambda_+\mu_{\nu}B_0\sin{\theta}\left[1-\left(1-\frac{\mu_{\nu}B_0\cos\theta}{4\lambda_+}\right)\frac{2V}{\lambda_+}+\frac{V^2}{\lambda_+^2}\right],
\end{equation} 
 and we get 
\begin{equation}
\Braket{+|\bm{\nabla}{+}}= \frac{i}{N^2B_0\sin{\theta}}\left(\frac{V}{2}+\mu_{\nu}B_0\cos{\theta}-\lambda_+\right)^2 \hat{\phi}.
\end{equation} 
The geometric phase factor acquired is then given by Eq.\eqref{eq:gamma}:
\begin{equation}
\gamma_+= i\int_{0}^{2\pi} B_0\sin{\theta}d\phi\Braket{+|\pmb{\nabla}{+}}= -\frac{2\pi}{N^2}\left(\frac{V}{2}+\mu_{\nu}B_0\cos{\theta}-\lambda_+\right)^2.
\end{equation} 
\\ 
We may re-write it as 
\begin{equation}\label{eq:longtnl}
\gamma_g(\xi, \theta)= \gamma_++2\pi= {\displaystyle 2\pi\left[1-\frac{1}{1+\left(\frac{\sin\theta}{\xi+\cos\theta-\sqrt{(\xi+\cos\theta)^2+\sin^2\theta})}\right)^2}\right]},
\end{equation} 
where $\xi= \frac{V}{2\mu_{\nu}B_0}$

	For the other case when neutrino is originally in the state $\Ket{-}$ we get the geometric phase $\gamma_-= -\gamma_+$.
	
	For the special case of neutrino propagating in a magnetic field precessing in the transverse direction i.e. $\theta= \pi/2$, we get 
\begin{equation}\label{eq:trans}
 \gamma_g^t(\xi)=  2\pi\left[1-\frac{1}{1+\left(\frac{1}{\xi-\sqrt{\xi^2+1}}\right)^2}\right].
\end{equation}

	It is particularly interesting to analyse the dependence of geometric phase on the energy of the neutrino. The energy dependence enters in Eq.\eqref{eq:longtnl} through the dependence of $V$ on the energy of the neutrino via the term $\Delta m^2/2E$. For the case of simple spin-precession, different helicity states have the same mass, i.e. $\Delta m^2=0$ and hence $\gamma_g$ in Eq.\eqref{eq:longtnl} is independent of energy, giving it a purely geometric character. For spin-flavor transitions $\Delta m^2 \neq 0$ , however as can be seen from \eqref{eq:longtnl}, for neutrinos with energy greater than 1 $GeV$ the geometric phase $\gamma_g$ is almost independent of energy.

\begin{figure}[h]
\begin{center}
\includegraphics{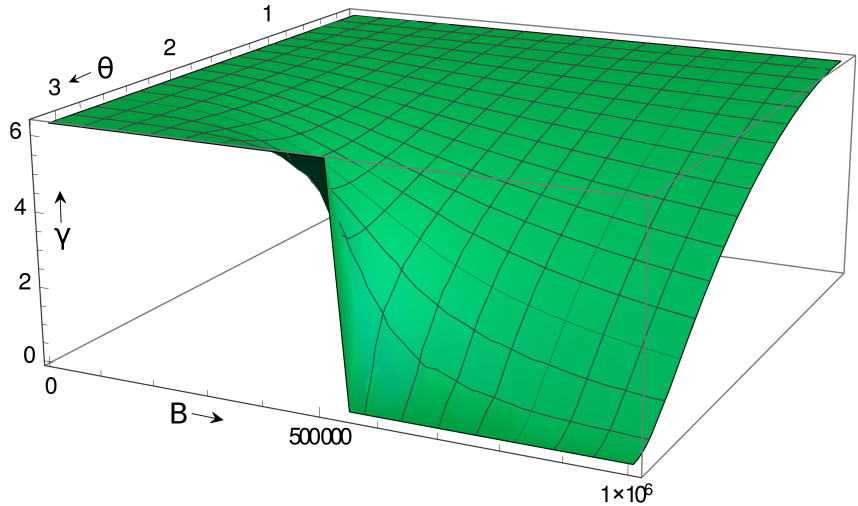}
\end{center}
\caption{Variation of geometric phase $\gamma_g$ can be seen here with magnetic field parameters $B_0(t)$ and $\theta(t)$.} 
\label{fig:phase}
\end{figure}

\section{Estimates}
Here we present the asymptotic behaviour of geometric phase and its variation with magnetic field parameters $\theta$ and $B$. We assume that neutrinos of energy $ E= 1 GeV$ are propagating in matter with density $n^{eff}= 10^{24} cm^{-3}$. For neutrino magnetic moment, we take the upper bound from GEMMA experiment\cite{gemma}:  $\mu_{\nu}= 2.9\times 10^{-11} \mu_B$ and $\Delta m^2= \Delta m^2_{sol}= 7.6\times10^{-5}eV^2$\cite{pdg}. The numbers taken here may vary depending on the source of neutrino and the medium in which neutrinos propagate, nonetheless they serve our purpose of getting rough estimates of the geometric phase obtained in Eq.\eqref{eq:longtnl}. For the above values of parameters we get $ \xi=5.24\times 10^5/B_0$.

	In the region of small fields ($B_0\sim$ few hundred gauss or less) $\xi\gg1$ and Eq.\eqref{eq:longtnl} gives  $\gamma_g \approx 2\pi$. Thus the effect of geometric phase will not be significant in this region. On the other hand in the region of extremely large fields ($B_0 > 10^{7}$ gauss) $\xi\ll1$ and we get from Eq.\eqref{eq:longtnl} $\gamma_g \approx \pi$. Thus if the magnetic field is sufficiently strong the neutrino state can develop significant geometric phase. 
	
	Using above value of $\xi$, in Figure \ref{fig:phase} we plot the geometric phase with magnetic field parameters $B_0(t)$ and $\theta(t)$. In Figure {\ref{fig:precession}, we show the geometric phase variation with $\log\xi$ for magnetic field precessing at a fixed angle about the neutrino direction for different precession angles.

\begin{figure}[t]
\begin{center}
\includegraphics{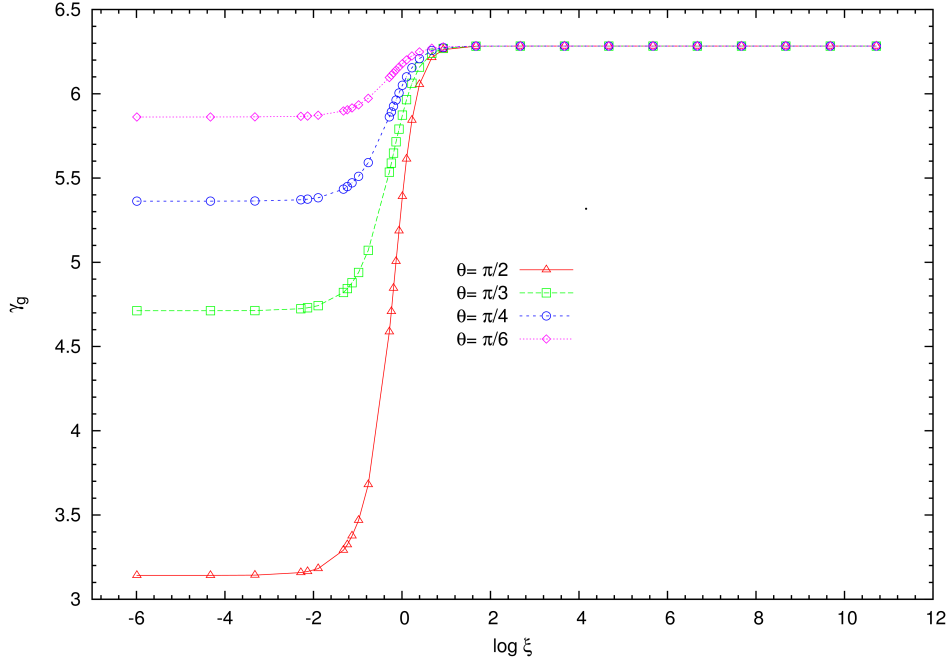}
\end{center}
\caption{The geometric phase $\gamma_g$ is seen to vary with $\log \xi$ from $\pi$ to $2\pi$ for magnetic field precessing about the direction of propagation of neutrino at different angles. The angle $\pi /2$ corresponds to field precession in a transverse plane with respect to neutrino propagation. Different points along the curve correspond to different astrophysical environments. Along the curve, from right to left, the magnetic field increases from about 0.1 $\mu G$ (galactic discs) to $10^{12} G$ (neutron stars or gamma ray bursts).} 
\label{fig:precession}
\end{figure}
	
	A particular case of interest is the transverse precession of the magnetic field, which contributes maximum to the geometric phase for a given field strength (for fields above $10^6 G$, see Figure \ref{fig:precession}). In this case geometric phase (Eq.\eqref{eq:trans}) assumes a particularly simple form, and we can invert this expression to obtain the factor $\xi$ in terms of the geometric phase $\gamma^t_g$:
\begin{equation}\label{eq:new1}
\xi= \frac{f_g-1}{\sqrt{2f_g-1}},
\end{equation}
where $f_g= \pi/\gamma_g^t$. Now, for the case of transverse field precession, the transition probablity $P_{\nu_L \rightarrow \nu_R}$ is given by Eq.\eqref{prob}, which we can write in terms of parameter $\xi$ as
\begin{equation}\label{new2}
P_{\nu_L \rightarrow \nu_R}(x)= \frac{1}{1+(\xi-\eta)^2}\sin^2[\mu_\nu B_0\sqrt{1+(\xi-\eta)^2}x],
\end{equation}
where $\eta= \dot{\phi}/2\mu_\nu B_0$. Substituting Eq.\eqref{eq:new1} in Eq.\eqref{new2}, we get the transition probability in terms of geometric phase as
\begin{equation}\label{eq:res}
P_{\nu_L \rightarrow \nu_R}(x)= \frac{1}{1+\left(\frac{f_g-1}{\sqrt{2f_g-1}}-\eta\right)^2}\sin^2\left[\mu_\nu B_0\sqrt{1+\left(\frac{f_g-1}{\sqrt{2f_g-1}}-\eta\right)^2}x\right].
\end{equation}

	Now Eq.\eqref{new2} gives the condition of resonance $\eta= \xi$, which using Eq.\eqref{eq:res} gives condition on $f_g$ as
\begin{equation}\label{res2}
 f_g^{res}= 1+\eta^2\pm\eta\sqrt{1+\eta^2}.
\end{equation} 

	For sufficiently strong magnetic fields ($\sim 10^7 G$ or more), using the upper bound on $\mu_\nu$ from GEMMA\cite{gemma}, we can see that under adiabatic conditions, $\eta=\dot{\phi}/2\mu_\nu B_0\ll1$, hence Eq.\eqref{res2} reduces to 
\begin{align}
f_g^{res}\approx& 1\pm\eta,\\
\mbox{or}, \quad \gamma_g^{t,{res}} \approx& \pi(1\mp\eta).
\end{align}

	Thus sufficiently strong magnetic fields may lead to resonance condition in the conversion probabality $P_{\nu_L \rightarrow \nu_R}$, where the geometric phase approaches the value $\pi$. At resonance, the transition probabilty $P_{\nu_L \rightarrow \nu_R}\approx \sin^2(\mu_\nu B_0x)$, has a characteristic length, $(\mu_\nu B_0)^{-1}$. In a region where $B\sim10^7 G$, for $\mu_\nu= 10 ^{-11} \mu_B$, this length is approximately 100 km.
\section{Conclusion}
In this Letter, we have calculated the  geometric phase when neutrinos propagate in magnetic fields in matter.  For strong enough magnetic fields ($\sim  10^7$ G or more), neutrinos can acquire a significant geometric phase which influences its spin precession and may lead to resonant spin-conversion $\nu_L\rightarrow\nu_R$ of neutrinos. Since the geometric phase appears directly in the transition probabilty $P({\nu_L \rightarrow \nu_R})$ (Eq.\eqref{eq:res}), it is, in principle, a directly observable quantity.

	In \cite{uhe} it was pointed out that the strong magnetic field prevailing in astrophysical environments such as Gamma ray bursts and Active galactic nuclei may cause the transition of active left handed neutrinos to right handed sterile neutrino. This would result in the reduction of the ultra high energy(UHE) neutrino flux. Our results compliment \cite{uhe} in this regard, as we have shown that sufficiently strong magnetic fields lead to a non-trivial geometric phase, which by inducing resonant transitions may make the UHE neutrinos unobservable. This mechanism might be important for experiments like IceCube\cite{icecube}, which detects UHE neutrinos coming from extragalactic sources. Eventually, these studies are expected to throw light on electromagnetic properties of neutrinos. 
	
%\section{Acknowledgments}
%We would like to take this opportunity to thank the Referee for valuable comments which helped us to improve the scope and content of the paper.

\end{document}